\documentclass[10pt, conference, letterpaper]{IEEEtran}
\makeatletter
\def\ps@headings{%
\def\@oddhead{\mbox{}\scriptsize\rightmark \hfil \thepage}%
\def\@evenhead{\scriptsize\thepage \hfil \leftmark\mbox{}}%
\def\@oddfoot{}%
\def\@evenfoot{}}
\makeatother 

\usepackage{color}
\usepackage{graphicx}               
\usepackage[cmex10]{amsmath}

\usepackage{url}
\usepackage{algorithm}
\usepackage{algpseudocode}
\usepackage{algorithmicx}

\begin{document}

\title{Real Time Vehicle Identification}

\author{\IEEEauthorblockN{Chandra Shekhar}\IEEEauthorblockA{Indian Institute of Technology, Bhubaneswar\\
cs13@iitbbs.ac.in}

\and \IEEEauthorblockN{Sudipta Saha}\IEEEauthorblockA{Indian Institute of Technology Bhubaneswar\\
sudipta@iitbbs.ac.in}

}

\maketitle

\IEEEpeerreviewmaketitle

\begin{abstract}
Identification of the vehicles passing over the roads is a very important component of an \textit{Intelligent Transportation System}. However, due to the presence of multiple vehicles together and their velocity, it gets hard to accurately identify and record them in real-time. Solutions based on Computer-vision use heavyweight equipment making them quiet inflexible, costly and hence unsuitable for wide-area coverage. Solutions based on RFID, although are lightweight and cost-effective, lack of fast and efficient communication protocol pertains to their inability to record multiple moving vehicles at the same time. We propose an IoT-assisted solution that leverages \textit{Synchronous-Transmission} based communication to bridge these gaps. Through extensive experiments we demonstrate that our strategy can consistently record upto an average of 40 vehicles running at speed range 30-90 Km/h with at least 97.5\% accuracy.
\end{abstract}
\vspace{-0.3cm}
\section{Introduction}
\label{sec:intro}
Real-time identification of the vehicles moving over the roads is one of the very significant components of an intelligent-transportation system \cite{ITS}. Several of the existing approaches to solve the problem exploit \textit{Computer-Vision} \cite{realtime-arya} along with \textit{OCR} techniques \cite{realtime-ocr} to read the \textit{Vehicle Registration Numbers} (VRN) directly from the number plates of vehicles. While such methods work, they require clear visibility of the number plate, installation of expensive equipment on the field, constant supply of power, constant cloud connectivity or heavy computation support. Such requirements make them highly inappropriate for widespread use. Moreover, these solutions also fail to appropriately scale with the number of vehicles passing over roads.

Active RFID-tags \cite{RFID-identification} have been also used to solve the problem. Under these solutions the vehicles are assumed to carry an active RFID-tag encoded with all the primary information of a vehicle. The RFID-readers are installed beside the road. During the passage of a vehicle, the readers scan the tags of the vehicles. These solutions are lightweight, cost-effective and hence, fit quite well for wide-spread use. However, due to lack of efficient communication and coordination protocols, they fail to achieve the goal effectively when many vehicles simultaneously pass by the readers \cite{RFID-SJ}. Classical multiple access problem as well as inadequate time for scanning all the vehicles equally pertain to such failures.

\begin{figure*}[htbp]
\begin{center}
\includegraphics[width=\textwidth]{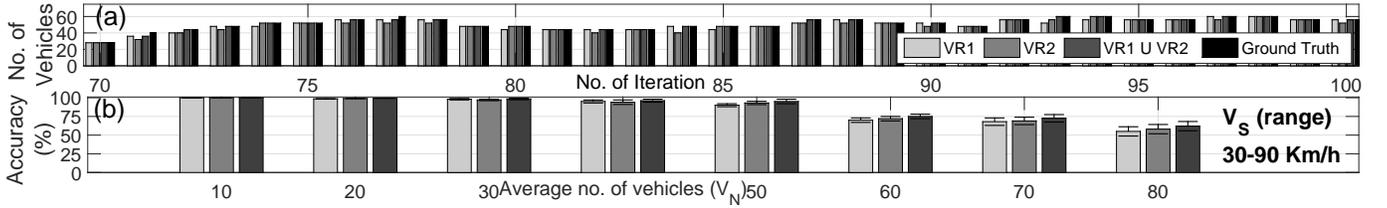}
\end{center}
\vspace{-0.5cm}
\caption{Accuracy of the proposed strategy in identifying vehicles.}
\vspace{-0.6cm}
\label{fig:result}
\end{figure*}

In this work we propose an IoT-based solution to solve this problem. The main challenge here is two-fold. \textit{First}, the number of vehicles passing by the static VRs at any point of time may be quite high. \textit{Second}, vehicles are supposed to be moving with a certain velocity, and hence the time available for a vehicle to become identified/recorded by a static recording utility installed as a road-side unit (here referred to as \textit{Vehicle Recorder} (VR)) would be quite low. Thus, in this work we perceive the vehicle recording job as a series of many-to-one communication instances between (many) vehicles and (one) VR under dynamic/hostile situation. 

Traditional \textit{Asynchronous-Transmission} (AT) based solutions for many-to-one data-sharing are naturally not suitable in this context. In recent works in IoT and WSN, \textit{Synchronous-Transmission} (ST) based strategies have gained quite good popularity because of their ability to achieve high reliability under very low-latency. Efficient operation of ST-based protocols have been also shown to sustain quite well even when the nodes are mobile \cite{mixer}. Hence, in the current work we step into exploiting ST-based strategies to design a protocol for solving the problem. Below we first provide a brief description of the design and subsequently provide our real-experiments and emulation based results to justify its effectiveness.

\vspace{-0.275cm}
\section{Design}

Each vehicle is supposed to carry an active entity, referred to as \textit{Electronic Number Plate} (ENP), holding the basic vehicle information, e.g., VRN etc. Both VR and ENP are equipped with 802.15.4 compatible RF transceiver. The ENPs act mostly in passive mode and answer the query asked by the VRs as needed. Our solution builds on top of  \textit{PacketSync} \cite{packetsync}, a ST-based based kernel, providing a basis for highly reliable low-latency many-to-one data-sharing between the VR and the ENPs. In brief, in PacketSync an initiator node sends a query message to a set of source nodes and provides a TDMA schedule using which the source nodes reply one by one following the principle of ST. In our setting a VR acts as the initiator and the ENPs installed at the vehicles act as the source nodes. 

PacketSync is designed considering a static setting. The TDMA schedule used in PacketSync assumes a defined number of source nodes having identifiers already known by the initiator. However, in the context of vehicle-recording the set of ENPs to be recorded by a VR is quite dynamic and undefined. VRN, being a unique number assigned to each vehicle is a natural choice to be used as the identifier for the ENPs. But, the whole set of the VRNs is too large to build a concrete TDMA schedule for use by PacketSync. It can be observed that, although the total possible VRN is very large, the number of ENPs present near a VR at a certain instance of time is supposed to be quite limited. Therefore, to solve the problem we take help of \textit{Hashing} as summarized below.

The query/probe message from the VR announces the total number of available slots ($S$), i.e., the maximum number of VRNs that can be accommodated in a single iteration of the protocol. Once it is received by the source ENPs, each of them first obtains a \textit{slot-number} by hashing its VRNs as per a predefined \textit{hash-function} and then responds to the query of the VR in that slot. Role of a suitable hash-function plays a very important role in this setting as even a single clash between two VRNs would make the corresponding ENPs to trying to push data at the same slot and would result in the missing of the data. 
We explicitly study the applicability of different hashing functions and converge to the well-known \textit{Mid Square Method}. \footnote{It is known to be highly efficient when the numerical values are bigger.}.

\textbf{Multiple VRs}: Fundamentally, no hashing strategy can guarantee 100\% collision free operation. However, use of ST helps in multiple ways to circumvent the problem. When two or more ENPs collide due to a clash by the hash-function, if the contending vehicles have different distances from the recording VR (which is highly likely), with high chance one of the contenders get recorded successfully by the virtue of the physical layer phenomena called \textit{Capture-Effect} (CE) \cite{capture-fm} which is an artifact of ST. We leverage this phenomena to maximize the accuracy by employing multiple VRs. In particular, we use two VRs in two opposite sides of the road. Each of the VRs executes one instance of PacketSync synchronously with the same hash-function. Due to different distances of the VRs from the vehicles (ENPs) this significantly enhances the chance of VRNs to get appropriately recorded in different VRs even when they both contend for the same slot due to hash-collision. Moreover, due to extremely fast operation of ST based PacketSync (of the order of ms), a vehicle gets multiple chances to interact with the VRs pairs during its contact period. 

\textbf{Full systems:} Several pairs of VRs are considered to be installed at different locations over the area under coverage. These VRs form a low-power wireless \textit{ad hoc}. The protocol Glossy \cite{glossy} is executed periodically to stay connected and synced. An iteration of the proposed strategy is composed by an instance of Glossy (completes in 20 ms, even for an area of 2000 x 2000 sq. meter) followed by an instance of the proposed protocol for the desired time (few hundreds of ms). 
\vspace{-0.2cm}
\section{Evaluation}
\label{sec:evaluation}

We implement the strategy in Contiki OS for TelosB. We experiment with the proposed strategy over local roads inside IIT Bhubaneswar campus with 10 vehicles (6 bikes and 4 cars) and one VR-pair with the vehicles passing the static VRs at different speed within a period of 5 mins. $S$ is set at 17. The process is repeated for 20 iterations where per iteration average accuracy observed is over 98.5\%. 

To understand the performance of the protocol under complex situation with more vehicles and higher speed variation we emulate the same in \textit{Cooja} under the MRM model. We set a network of 5 VR-pairs covering a road of length 200 meters. Glossy is executed with a period of 512 ms. Cooja is appropriately modified with the support for run-time mobility of the vehicles. Both the speed ($V_S$) and the number ($V_N$) of vehicles passing over the road are varied over a wide range. $S$ is set at 71. Each experiment is repeated at least 1000 iterations. 

Fig.~\ref{fig:result}(a) shows the number of vehicles detected per iteration by each of the VRs in one of the VR-pairs, as well as their union ($A_u$) in one of the experiments for $V_N$ = 50 and $V_S$ (range) = 30-90 km/h. It also shows the Ground-Truth (GT) obtained from Cooja by observing the number of vehicles present within the area covered by the transmission/reception range of the VR-pair. Fig.~\ref{fig:result}(b) shows the recording accuracy for different $V_N$. It can be observed that with $V_S$ (range) = 30-90km/h, for even upto $V_N$ = 40, $A_u$ stays above 97.5\%. However, it starts degrading slowly with the increase in $V_N$.

\vspace{-0.2cm}
\section{Conclusion}

Real-time recording of the identities of the vehicle passing over the roads is a very important module. It an extremely challenging job considering both the speed and the number of the vehicles. We propose a lightweight flexible ST-based protocol to solve the problem. The effectiveness of the same is demonstrated through extensive experiments with real-devices and emulated settings.

\bibliographystyle{abbrv}

\end{document}